\documentclass[epj]{svjour}
\usepackage{graphics,lscape,graphicx,epsfig,amsmath,amssymb,epstopdf}
\usepackage{esint,graphicx}
\usepackage{color}
\usepackage{comment}
\usepackage[cal=boondox]{mathalfa}
\usepackage{tabu}
\allowdisplaybreaks
\newcommand{\be}{\begin{equation}}
\newcommand{\ee}{\end{equation}}
\newcommand{\bea}{\begin{eqnarray}}
\newcommand{\eea}{\end{eqnarray}}
\newcommand{\ba}{\begin{eqnarray*}}
\newcommand{\ea}{\end{eqnarray*}}
\newcommand{\dagga}{{\phantom{\dagger}}}
\newcommand{\bR}{\mathbf{R}}
\newcommand{\bQ}{\mathbf{Q}}
\newcommand{\bK}{\mathbf{K}}
\newcommand{\bq}{\mathbf{q}}

\newcommand{\bk}{\mathbf{k}}

\newcommand{\bp}{\mathbf{p}}

\newcommand{\bx}{\mathbf{x}}

\newcommand{\br}{\mathbf{r}}
\newcommand{\bG}{\mathbf{G}}

\newcommand{\dis}{\displaystyle}

\newcommand{\fract}[2]{\frac{\dis #1}{\dis #2}}

\newcommand{\eqn}[1]{(\ref{#1})}

\newcommand{\bsigma}{\boldsymbol{\sigma}}
\newcommand{\bGamma}{{\boldsymbol{\Gamma}}}
\newenvironment{eqs}%
{\begin{equation} \begin{aligned}}%
{\end{aligned} \end{equation} }
\newcommand{\beal}{\begin{eqs}}
\newcommand{\eal}{\end{eqs}}
\newcommand{\bw}{\begin{widetext}}
\newcommand{\ew}{\end{widetext}}
\newcommand{\esp}[1]{\text{e}^{#1}}

\newcommand{\bM}{\mathbf{M}}
\newcommand{\bd}[1]{{\boldsymbol{#1}}}
\newcommand{\bC}[1]{\text{C}_{#1}}
\newcommand{\mo}{{moir\'e~}}
\newcommand{\bA}{\mathbf{A}}
\newcommand{\bB}{\mathbf{B}}
\newcommand{\due}{{(2)}}
\newcommand{\uno}{{(1)}}

\newcommand{\apa}{{(a)}}
\newcommand{\apb}{{(b)}}
\begin{document}
\title{
Jahn-Teller coupling to \mo phonons in the continuum model formalism for 
small angle twisted bilayer graphene}
\author{Mattia Angeli\inst{1} \and Michele Fabrizio\inst{1}}  
\institute{International School for
  Advanced Studies (SISSA), Via Bonomea
  265, I-34136 Trieste, Italy}
\date{Received: date / Revised version: date}
\abstract{
We show how to include the Jahn-Teller coupling of \mo phonons to the electrons in the continuum model formalism which describes small angle twisted bilayer graphene. These phonons, which strongly couple to the valley degree of freedom, are able to open gaps at most  integer fillings of the four flat bands around the charge neutrality point.
Moreover, we derive the full quantum mechanical expression of the electron-phonon Hamiltonian, which may allow accessing phenomena such as the phonon-mediated superconductivity and the dynamical Jahn-Teller effect.
\PACS{
      {PACS-key}{discribing text of that key}   \and
      {PACS-key}{discribing text of that key}
     } % end of PACS codes
} %end of abstract
\maketitle

\section{Introduction}

The discovery of superconductivity first in small angle twisted bilayer graphene (tBLG) \cite{Herrero-1,Herrero-2,Yankowitz,Efetov}, and later in 
trilayer \cite{Wang_arxiv} and double bilayer graphene \cite{Kim_tdbg}, has 
stimulated an intense theoretical and experimental research activity.
In these systems, the twist angle tunes a peculiar interference within a large set of energy bands, compressing energy levels to form a set of extremely narrow bands around charge neutrality \cite{Fabrizio_PRX,Fabrizio_PRB,Kaxiras-2,Vishvanath_WO}. In twisted bilayer graphene, these flat bands (FBs) have a bandwidth of the order of $\approx 10-20$ meV, and are isolated in energy by single particle band gaps of the order of $30-50$ meV. Superconductivity 
is observed upon doping such narrow bands, often surrounding insulating states at 
fractional fillings that contradict the metallic behaviour predicted by band structure calculations.
The observed phenomenology of these insulating states,
which turn metallic above a threshold Zeeman splitting or above a critical temperature, suggests that they might arise from a weak-coupling Stoner or CDW band instability driven by electron-electron and/or electron-phonon interactions, rather than from the Mott's localisation phenomenon in presence of strong correlations.
This is further supported by noting that the effective \textit{on-site} Coulomb repulsion $U$ 
must be identified with the charging energy of the supercell, which can be as large as tens of nanometers at small angles, projected onto the flat bands.
If screening effects due to the gates and to the other bands are taken into account,
the actual value of $U$ is estimated of the order of few meVs, suggesting that tBLG might not be more correlated than a single graphene sheet \cite{U-graphene}.
On the contrary, there are evidences that the coupling to the lattice is instead anomalously  large if compared with the FBs bandwidth. For instance, ab-initio DFT-based calculations fail to predict well defined FBs separated from other bands, unless atomic positions are allowed to relax \cite{Nam_Koshino_PRB,Kaxiras,Fabrizio_PRB,Procolo,Kaxiras-2,Choi}, in which case gaps open that are larger than the FBs bandwidth. Further evidences supporting a
sizeable electron-phonon coupling come from transport properties \cite{Polshyn,Sarma,Vignale}, but also from direct electronic structure calculations.\\
Specifically, in Ref.~\cite{Fabrizio_PRX} it was shown that a pair of optical phonon modes are rather strongly coupled to the FBs, and thus might play an important role in the 
physics of tBLG. These modes, which have a long wavelength modulation on the same moir\'e scale, have been dubbed as 'moir\'e phonons', and recently observed experimentally \cite{Jorio}. 
%Their coupling to the electronic degree of freedom realises an $e \otimes E$ Jahn-Teller model, which is able to efficiently lift the degeneracies in the band structure by appropriately breaking symmetries of the system, thus opening gaps at fractional fillings of the flat bands.The large electron phonon coupling constants involved in this process, suggests the possibility that these modes are also playing a role in mediating superconductivity in a BCS framework.
However, the large number of atoms contained in the small angle unit cell of twisted bilayer graphene (more than $11000$), make any calculation, more involved than a simple thigh-binding one, rather tough, if not computationally impossible.
In this paper we try to cope with such problem by implementing the effect of these phonon modes on the band structure in the less computationally demanding continuum model of Ref.~\cite{MacDonald}. This method can serve as a suitable starting point for BCS \cite{MacDonald_phBCS,Bernevig_phBCS}, Hartree-Fock \cite{Vishvanath_HF,MacDonald_HF,guinea_HF,zhang_HF} and many other calculations, which may involve both phonons and correlations.
The work is organized as follows. In section \ref{CM} we derive the Bistritzer and MacDonald continuum model for twisted bilayer graphene. In section 
\ref{phonon} we implement in the continuum model formalism the effect of a static atomic displacement. By using lattice deformation fields which are similar to the Jahn-Teller moir\'e phonon modes of ref. \cite{Fabrizio_PRX}, we show how the band structure and the density of states of the system evolves as a function of the lattice distortion intensity.
Finally,  \ref{conclusion} is devoted  to concluding remarks.

\section{Continuum Model Hamiltonian} 
\label{CM}
We start by introducing the Bistritzer-MacDonald continuum model for twisted bilayer graphene \cite{MacDonald}, and recall that the single layer Dirac Hamiltonian 
is, around the $\bK$ and $\bK'=-\bK$ valleys, respectively, 
\beal
\hat H_{\bk\sim \bK} &\equiv \hat H^{+1}_\bk = -v\,\Big(\bk-\bK\Big)\cdot
\Big(\sigma_x,\sigma_y\Big)\,,\\
\hat H_{\bk\sim -\bK} &\equiv \hat H^{-1}_\bk = -v\,\Big(\bk+\bK\Big)\cdot
\Big(-\sigma_x,\sigma_y\Big)\,,
\eal
namely
\beal
\hat H^\zeta_\bk &= -v\,\Big(\bk-\zeta\bK\Big)\cdot\Big(\zeta\,\sigma_x,\sigma_y\Big)\,,
\eal
with $\zeta=\pm 1$ the valley index, and where the Pauli matrices $\sigma_a$ 
act on the two component wavefunctions, each component referring to one of the 
two sites per unit cell that form the honeycomb lattice, 
which we shall hereafter denote as sublattice $A$ and $B$.\\
\begin{figure}[thb]
\centerline{\includegraphics[width=0.7\textwidth]{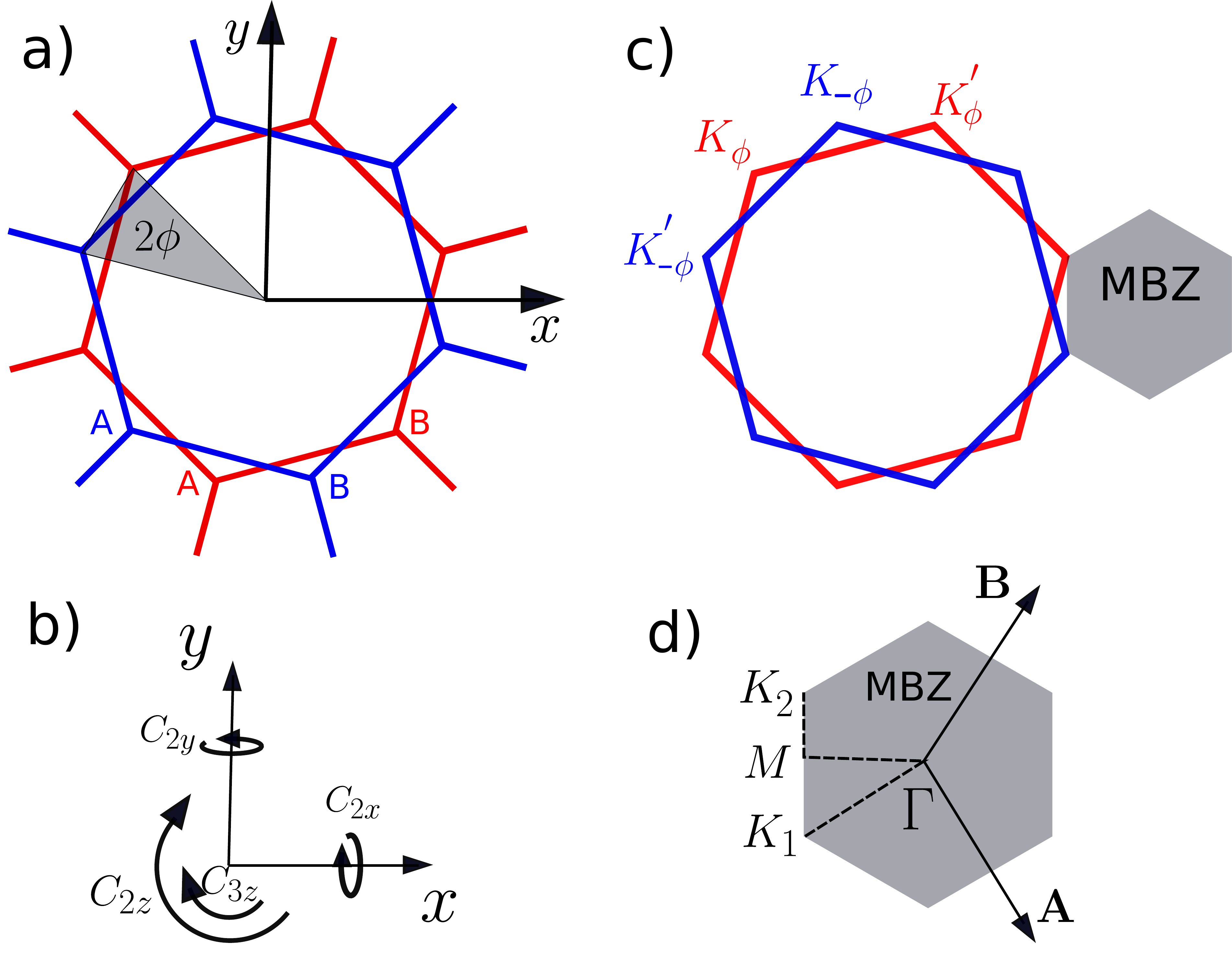}
}
\caption{a) Two graphene layers twisted with respect to the center of two overlapping hexagons by an angle $2\phi$. b) Symmetries of twisted bilayer graphene which are those of the group $\text{D}_6$. Two- ($180^\circ$) and three-fold ($120^\circ$) rotations with respect to the $z$-axis denoted as $C_{2z}$ and $C_{3z}$ as well as two-fold rotations  with respect to the $x$ and $y$-axis denoted as $C_{2y}$ and $C_{2x}$. c) The monolayer Brillouin zones that are folded in the mini-Brillouin zone (MBZ) of the twisted heterostructure.d) Mini-Brillouin zone of twisted bilayer graphene with reciprocal lattice vectors $\bA$ and $\bB$. The high-symmetry points and paths are highlighted.  }
\label{fig1}
\end{figure}\\
Our analysis must start by defining the specific twisted bilayer graphene we shall investigate, and by setting some conventional notations. 
We assume that the twisted bilayer is obtained 
by rotating two AA-stacked layers ($i=1,2$) by an opposite angle $\pm \phi$ with respect to the center of two overlapping hexagons, see Fig.~\ref{fig1}(a), where 
$\tan\phi = 1/\big[(2k+1)\sqrt{3}\big]$, with large positive integer $k$. With this choice, the moir\'e pattern forms a superlattice, which is still honeycomb and endowed by a $\text{D}_6$ 
space group symmetry \cite{Fabrizio_PRB} that is generated by the three-fold rotation 
$\bC{3z}$ around the $z$-axis perpendicular to the bilayer, and the two-fold rotations 
around the in-plane $x$ and $y$ axes, $\bC{2x}$ and $\bC{2y}$, respectively. \\
The corresponding mini-Brillouin zone (MBZ) has reciprocal lattice vectors 
\beal
\bA&=\bG_2^\due-\bG_2^\uno\,,& \bB&=\bG_1^\uno-\bG_1^\due\,.
\label{AB-G}
\eal
where $\bd{G}^{(1)}_{1/2}=\mathcal{R}_{+\phi}\big(\bd{G}_{1/2}\big)$ 
and $\bd{G}^{(2)}_{1/2}=\mathcal{R}_{-\phi}\big(\bd{G}_{1/2}\big)$ are the reciprocal lattice vectors of each layer after the twist, with $\mathcal{R}_\phi$ the rotation operator by an  angle $\phi$.\\
The Dirac nodes of each monolayer are, correspondingly, $
\mathcal{R}_{\pm\phi}(\bK)\equiv \bK_{\pm \phi}$ for the valley we  
shall denote as $\zeta=+1$, and $\mathcal{R}_{\pm\phi}(\bK')\equiv \bK'_{\pm \phi}$ for the other valley, $\zeta=-1$. With our choices, $\bK_{+\phi}$ and $\bK'_{-\phi}$ 
fold into the same point $\bK_2$ of the MBZ, as well as $\bK_{-\phi}$ and $\bK'_{+\phi}$
into the point $\bK_1$, see Fig.~\ref{fig1}. \\
We introduce the (real) Wannier functions derived by the $p_z$ orbital 
of each carbon atom:
\beal
\phi_{1\, \alpha\bR^{(1)}}(\br) &\equiv \phi\bigg(\br-\bR^{(1)}-\br^{(1)}_\alpha 
-\fract{\bd{d}_\perp}{2}\bigg)\,,\\ 
\phi_{2\,\alpha\bR^{(2)}}(\br) &\equiv \phi \bigg(\br-\bR^{(2)}-\br^{(2)}_\alpha 
+\fract{\bd{d}_\perp}{2}\bigg)\,,
\eal
where $\bd{d}_\perp = (0,0,d)$, with $d$ the interlayer distance, 
$\bR^{(i)}$ label the positions of the unit cells in layer $i=1,2$, while 
$\br^{(i)}_\alpha$ the coordinates with respect to $\bR^{(i)}$ of the two sites within each unit cell, $\alpha=A,B$ denoting the two sublattices. From the Wannier functions we 
build the Bloch functions
\beal 
\psi_{1\,\alpha\bk}(\br) &= \fract{1}{\sqrt{N}}\,\sum_{\bR^\uno}\, 
\esp{-i\bk\cdot\big(\bR^{(1)}+\br^{(1)}_\alpha\big)}\;
\phi_{1\, \alpha\bR^{(1)}}(\br)\,,\\
\psi_{2\,\alpha\bk}(\br) &= \fract{1}{\sqrt{N}}\,\sum_{\bR^{(2)}}\, 
\esp{-i\bk\cdot\big(\bR^{(2)}+\br^{(2)}_\alpha\big)}\;\phi_{2\,\alpha\bR^{(2)}}(\br)\,.
\label{k-WO}
\eal
Conventionally, one assumes the two-center approximation \cite{MacDonald}, so that, if $V_\perp(\br)$ is the interlayer potential, then the interlayer hopping 
\beal
\int d\br\, \phi_{1\, \alpha\bR^\due}(\br)\;V_\perp(\br)\;
\phi_{2\,\beta\bR^\uno}(\br) 
&\simeq T_\perp \Big(\bR^\due+\br^\due_\alpha - \bR^\uno-\br^\uno_\beta\Big)\,,
\label{T(r)}
\eal 
depends only on the distance between the centers of the two Wannier orbitals. We define 
$T_\perp(\bq)$, the Fourier transform of $T_\perp(\br)$:
\be
T_\perp(\br) = \fract{1}{N}\,\sum_\bq\, \esp{i\bq\cdot\br}\;T_\perp(\bq)\,,\label{T(q)}
\ee
where $\br$ and $\bq$ are vectors in the $x$-$y$ plane. Hereafter, all momenta are assumed 
also to lie in the $x$-$y$ plane.\\   
The interlayer hopping between an electron in layer 1 with momentum $\bp$ and one in layer 2 with momentum $\bk$ is in general a matrix $\hat T_{\bk\bp}$, with 
elements $T^{\alpha\beta}_{\bk\bp}$, $\alpha,\beta=A,B$, which, through 
equations \eqn{k-WO}, \eqn{T(r)} and \eqn{T(q)}, reads explicitly
\beal
T^{\alpha\beta}_{\bk\bp} &= \fract{1}{N}\,\sum_{\bR^\due\bR^\uno}\, 
\esp{-i\bk\cdot\big(\bR^\due+\br^\due_\alpha\big)}\;
\esp{i\bp\cdot\big(\bR^\uno+\br^\uno_\beta\big)}\;
T_\perp\big(\bR^\due+\br^\due_\alpha - \bR^\uno-\br^\uno_\beta\big)\\
&= \sum_\bq\, T_\perp(-\bq)\;\; \fract{1}{N^2}\,\sum_{\bR^\due\bR^\uno}\, 
\esp{-i\big(\bk+\bq\big)\cdot\big(\bR^\due+\br^\due_\alpha\big)}\;\esp{i\big(\bp+\bq\big)\cdot\big(\bR^\uno+\br^\uno_\beta\big)}\\
&= \sum_\bq\, T_\perp(-\bq)\;\sum_{\bG^\due \bG^\uno}\,
\delta_{\bk+\bq\,,\,-\bG^\due}\,\delta_{\bp+\bq\,,\,-\bG^\uno}\,
\esp{i\bG^\due\cdot\br^\due_\alpha}\;\esp{-i\bG^\uno\cdot\br^\uno_\beta}\\
&= \sum_{\bG^\due \bG^\uno}\,T_\perp\Big(\bk+\bG^\due\Big)\;
\delta_{\bk+\bG^\due\,,\,\bp+\bG^\uno}\;\esp{i\bG^\due\cdot\br^\due_\alpha}\;\esp{-i\bG^\uno\cdot\br^\uno_\beta}\;.\label{T-inter}
\eal
Since we are interested in the low energy physics, $\bk$ and $\bp$ must be close 
to the corresponding Dirac points, namely $\bK_\phi$ and $\bK'_\phi$ for $\bp$ in layer 1, 
while $\bK_{-\phi}$ and $\bK'_{-\phi}$ for $\bk$ in layer 2. Therefore, $\hat T_{\bk\bp}$ can in principle couple to each other 
states of different layers within the same valley, or between opposite valleys. Since $T_\perp(\bq)$ decays exponentially with $q=|\bq|$ \cite{MacDonald}, the leading terms are those with the least possible $\big|\bk+\bG^\due\big|$ compatible with momentum conservation $\bk+\bG^\due=\bp+\bG^\uno$. At small twist angle $\phi$, only the intra-valley matrix elements, $\bp\sim \bk$, are sizeable, while the inter-valley ones are negligibly small, despite opposite valleys of different layers fold into the same point of the MBZ.  
For instance, if $\bp\simeq \bK_{+\phi}$ and $\bk\simeq \bK'_{-\phi}$,   
momentum conservation requires very large 
$\bG^\uno = (2k+1)\,\big(\bG_2^\uno-\bG_1^\uno\big)$
and $\bG^\due =(2k+1)\,\big(\bG_2^\due-\bG_1^\due\big)$, thus an exponentially small 
$T_\perp\big(\bk+\bG^\due\big)$. The effective decoupling between the two valleys 
implies that the number of electrons within each valley is to high accuracy a conserved quantity, thus an emergent valley $U_v(1)$ symmetry \cite{MacDonald,Senthil_PRX} that 
causes accidental band degeneracies along high-symmetry lines in the MBZ \cite{Senthil_PRX,Fabrizio_PRX}. \\

\noindent
We can therefore just consider the intra-valley inter-layer scattering processes. We  
start with valley $\zeta=+1$, and thus require that 
$\bk$ is close to $\bK_{-\phi}=\bK_1$ and $\bp$ close to $\bK_{+\phi}=\bK_2$, 
see Fig.~\ref{fig1}(d). Since the modulus 
of $\bk\sim \bK_1$ is invariant under $C_{3z}$ rotations, where 
$C_{3z}\big(\bK_1\big) = \bK_1-\bG^\due_1$ and $C_{3z}^2\big(\bK_1\big) = \bK_1-\bG^\due_2$, 
maximisation of $T_\perp\big(\bk+\bG^\due\big)$ compatibly with 
momentum conservation leads to the following conditions, see Eq.~\eqn{AB-G}, 
\beal
\bp &= \bk\,,\\ 
\bp &= \bk - \bG^{(2)}_1 + \bG^{(1)}_1 =\bk +\bd{B}\,,\\
\bp &= \bk - \bG^{(2)}_2 + \bG^{(1)}_2 = \bk - \bd{A}\,.\label{cond-k}
\eal
Upon defining $T(\bk)\equiv t_\perp$, 
and using Eq.~\eqn{cond-k} to evaluate the phase factors in \eqn{T-inter}, 
we finally obtain 
\beal
\hat T_{\bk\bp}^{\zeta=+1} = \delta_{\bp,\bk}\;\hat T_1 +
\delta_{\bp\,,\,\bk+\bd{B}}\;
\hat T_2 + \delta_{\bp\,,\,\bk-\bd{A}}\;\hat T_3\;,\label{T-kp}
\eal
where we explicitly indicate the valley index $\zeta$, and   
\beal
\hat T_1 &= t_\perp\,
\begin{pmatrix}
1 & 1\\
1 & 1
\end{pmatrix}\;,&
\hat T_2 &=t_\perp\,\begin{pmatrix}
1& \omega^*
\\
\omega &1
\end{pmatrix}\,,
&
\hat T_3 &= t_\perp\,\begin{pmatrix}
1 & \omega
\\
\omega^* & 1
\end{pmatrix}\,,
\label{matrix}
\eal
with $\omega=\esp{2\pi i/3}$. \\

\noindent
We now focus on the other valley, $\zeta=-1$, and take $\bk$ close to 
$\bK'_{-\phi}=-\bK_2$, and $\bp$ to $\bK'_{\phi}=-\bK_1$, see Fig.~\ref{fig1}(d). In this case Eq.~\eqn{cond-k} is replaced by 
\beal
\bp &= \bk\,,\\ 
\bp &= \bk -\bd{B}\,,\\
\bp &= \bk + \bd{A}\,,
\eal
and
\beal
\hat T_{\bk\bp}^{\zeta=-1} = \delta_{\bp,\bk}\;\hat T_1^* +
\delta_{\bp\,,\,\bk-\bd{B}}\;
\hat T_2^* + \delta_{\bp\,,\,\bk+\bd{A}}\;\hat T_3^*\;.\label{T-kp-next valley}
\eal\\

\noindent
Let us briefly discuss how one can take into account lattice relaxation, which is known to 
shrinks the energetically unfavourable AA regions enlarging the Bernal stacked triangular domains in the moir\'e pattern \cite{Yoo_NatMat,Guinea_relax,Fabrizio_PRB,Yazyev,Procolo,Kaxiras,Nam_Koshino_PRB}.
As a consequence, the  inter- and intra-sublattice hopping processes acquire different amplitudes, which is taken into account by modifying the operators 
$\hat T_i$ in Eq.~\eqn{matrix} according to 
\beal
\label{relax}
\hat T_1 &\to T_1(u,u')= 
\begin{pmatrix}
u & u'\\
u' & u
\end{pmatrix} = u\,\sigma_0 + u'\,\sigma_x\;,\\
\hat T_2 &\to \hat T_2(u,u')=\begin{pmatrix}
u& u'\,\omega^*
\\
u'\,\omega &u
\end{pmatrix}= u\,\sigma_0 + u'\Big(
\cos\fract{2\pi}{3}\,\sigma_x + \sin\fract{2\pi}{3}\,\sigma_y\Big)
\,,\\
\hat T_3 &\to \hat T_3(u,u')= \begin{pmatrix}
u & u'\,\omega
\\
u'\,\omega^* & u
\end{pmatrix}= u\,\sigma_0 + u'\Big(
\cos\fract{2\pi}{3}\,\sigma_x -\sin\fract{2\pi}{3}\,\sigma_y\Big)\,,
\eal
with $u$ generally smaller than $u'$.\\ 
 
\noindent
We conclude by showing how this formalism allows recovering the untwisted case, where $\bG^{(1)}_{1/2}=\bG^{(2)}_{1/2}$, so that, through Eq.~\eqn{AB-G}, $\bd{A}=\bd{B}=\bd{0}$,  
and therefore 
\be
\hat T_{\bk\bp} \;\underset{\phi\to 0}{\xrightarrow{\hspace*{.8cm}}}\;  \delta_{\bp,\bk}\;\Big(\hat T_1
+\hat T_2+\hat T_3\Big)  = 3t_\perp\,\delta_{\bp,\bk}\,
\begin{pmatrix}
1 & 0\\
0 & 1
\end{pmatrix}\,,
\ee
which is what one would expect from an AA stacked bilayer.

\subsection{A more convenient representation}
\label{A more convenient representation}
For our purposes, it is actually more convenient to use the alternative representation 
of the Hamiltonian derived in Ref.~\cite{Bernevig_PRL}.
We translate $\bK'_{\phi}=-\bK_1$ so that it falls on $\bK_{-\phi}=\bK_1$, and 
similarly $\bK'_{-\phi}=-\bK_2$ on $\bK_{\phi}=\bK_2$, see Fig.~\ref{fig1}(d). 
This implies that the diagonal parts of the Hamiltonian $\hat H^{(i)}_\zeta(\bk)$, where $i=1,2$ is 
the layer index and $\zeta=\pm 1$ the valley one, become simply 
\beal 
\hat H^\due_{+1}(\bk) &= - v\, \Big(\bk-\bK_1\Big)\cdot\Big(\sigma_x\,,\,\sigma_y\Big)\,,\\
\hat H^\uno_{-1}(\bk) &= - v\, \Big(\bk-\bK_1\Big)\cdot\Big(-\sigma_x\,,\,\sigma_y\Big)
= v\, \Big(\bk-\bK_1\Big)\cdot\bsigma^T\,,\\
\hat H^\uno_{+1}(\bk) &= - v\, \Big(\bk-\bK_2\Big)\cdot\Big(\sigma_x\,,\,\sigma_y\Big)\,,\\
\hat H^\due_{-1}(\bk) &= - v\, \Big(\bk-\bK_2\Big)\cdot\Big(-\sigma_x\,,\,\sigma_y\Big)
=v\, \Big(\bk-\bK_2\Big)\cdot\bsigma^T
\,.
\eal
\begin{figure}
\vspace{-1cm}
\centerline{\includegraphics[width=0.7\textwidth]{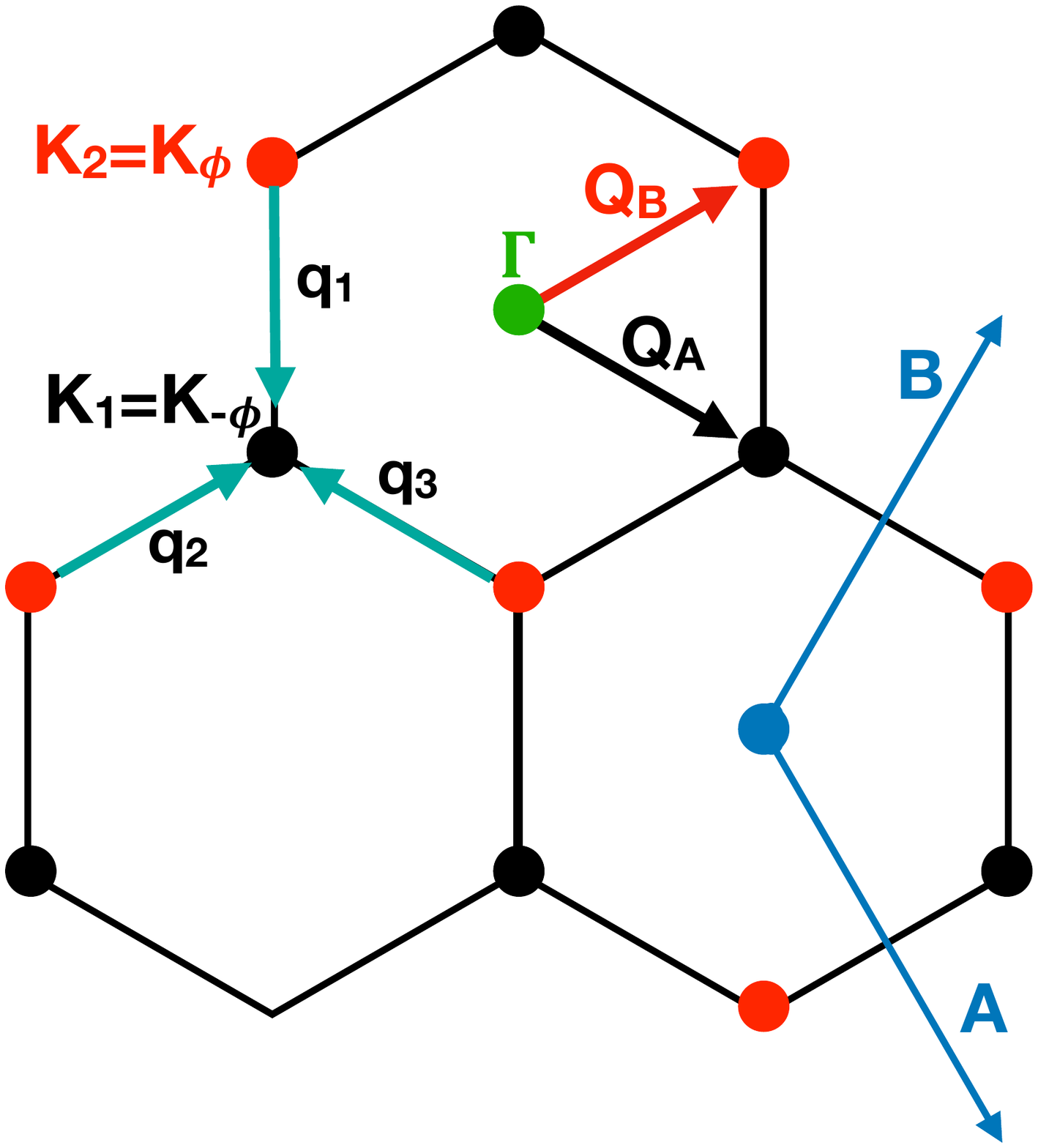}}
\vspace{-.5cm}
\caption{The lattice in momentum space of Ref.~\cite{Bernevig_PRL}. The $\bQ_{A}$ and $\bQ_{B}$ vectors span the lattice formed by the $\bK$ points of the two twisted monolayer Brillouin zones.}
\label{convention}
\end{figure}
Following Ref.~\cite{Bernevig_PRL}, we define a set of vectors 
\beal
\bQ &= \Big\{\bQ_A\,,\,\bQ_B\Big\}=
\begin{cases}
\bQ_A = \bK_1 + n\bA + m\bB\,,\\
\bQ_B = \bK_2 + n\bA + m\bB\,, 
\end{cases}\label{def:Q-vectors}
\eal
which span the vertices of the MBZs, where $\bQ_A$, black circles in Fig.~\ref{convention}, 
correspond to valley $\zeta=+1$ in layer 2 and valley $\zeta=-1$ in layer 1, 
while $\bQ_B$, red circles in Fig.~\ref{convention}, 
correspond to valley $\zeta=+1$ in layer 1 and valley $\zeta=-1$ in layer 2. In addition    
we define 
\beal
\bq_1 &= \bK_{1}-\bK_{2}\,,&
\bq_2 &= \bq_1+\bB\,,&
\bq_3 &= \bq_1-\bA\,. \label{def:q}
\eal
Next, we redefine the momenta for layers 1 and 2 as, respectively,
\beal
\bp -\bK_{2}&\to \bk' - \bQ_B\,,&
\bk -\bK_{1}&\to \bk' - \bQ_A\,,
\eal
thus
\beal
\bk &= \bk'+\bK_{1}-\bQ_A\,,&
\bp &= \bk'+\bK_{2}-\bQ_B\,,
\eal
so that the selection rules transforms into 
\beal
\bp &= \bk \;\Rightarrow\; \bQ_A-\bQ_B = \bq_1
\;\Rightarrow\; \bQ_B = \bQ_A -\bq_1
\,,\\
\bp &= \bk +\bB\;\Rightarrow\; \bQ_A-\bQ_B = \bq_1+\bB
\;\Rightarrow\; \bQ_B = \bQ_A -\bq_2\,,\\
\bp &= \bk -\bA\;\Rightarrow\; \bQ_A-\bQ_B = \bq_1-\bA
\;\Rightarrow\; \bQ_B = \bQ_A -\bq_3\,.
\label{sel_rules}
\eal 
With those definitions, and denoting the conserved momentum $\bk'$ as $\bk$, the Hamiltonian of valley $\zeta$ now reads
\beal
\hat{H}^\zeta_{\bQ \bQ'}(\bk) 
&= \delta_{\bQ\bQ'}\,v\,\zeta\,\big(\bk-\bQ\big)\cdot\big(\sigma_x,\zeta\,\sigma_y\big) \\
&\quad + \sum_{i=1}^3\,\Big(\delta_{\bQ'-\bQ,\bq_i}
+ \delta_{\bQ-\bQ',\bq_i}\Big)\,\hat T^\zeta_i(u,u')\,,
\eal
where 
\beal
\hat T^\zeta_1(u,u') &= u\,\sigma_0 + u'\,\sigma_x\,,\\
\hat T^\zeta_2(u,u') &= u\,\sigma_0 + u'\,\Big(\cos\frac{2\pi}{3}\;\sigma_x
+\zeta\,\sin\frac{2\pi}{3}\;\sigma_y\Big)\,,\\
\hat T^\zeta_3(u,u') &= u\,\sigma_0 + u'\,\Big(\cos\frac{2\pi}{3}\;\sigma_x
-\zeta\,\sin\frac{2\pi}{3}\;\sigma_y\Big) = \sigma_x\;\hat T^\zeta_2(u,u')\;\sigma_x
\,.
\eal
 In particular, 
\beal
\hat T^{-\zeta}_i(u,u') &= \sigma_x\,\hat T^{\zeta}_i(u,u')\,\sigma_x\,.
\eal
One can further simplify the notation introducing the Pauli matrices $\tau_a$, $a=0,x,y,z$, 
with $\tau_0$ the identity, that act in the valley subspace, and thus write 
\beal
\hat{H}_{\bQ \bQ'}(\bk) 
&= \delta_{\bQ\bQ'}\,v\,\tau_z\,\big(\bk-\bQ\big)\cdot\,\Omega\,\bsigma\,\Omega\\
&\qquad + \tau_0\,\sum_{i=1}^3\,\Big(\delta_{\bQ'-\bQ,\bq_i}
+ \delta_{\bQ-\bQ',\bq_i}\Big)\;\Omega\,\hat T_i(u,u')\,\Omega\,,
\eal
where $\hat T_i(u,u') \equiv \hat T^{+1}_i(u,u')$, and $\Omega$ is the real unitary operator 
\beal
\Omega &= \sigma_x\;\fract{1-\tau_z}{2} + \sigma_0\;\fract{1+\tau_z}{2}
= \sigma_x\,\text{P}_{\zeta=-1} + \sigma_0\,\text{P}_{\zeta=+1}\;,
\label{U_oper}
\eal
being P$_\zeta$ the projector onto valley $\zeta$, which actually 
interchanges sublattice $A$ with $B$ in the valley $\zeta=-1$. Applying the unitary operator 
$\Omega$ we thus obtain 
\beal
\Omega\,\hat{H}_{\bQ \bQ'}(\bk)\,\Omega 
&\to \hat{H}_{\bQ \bQ'}(\bk)
= \delta_{\bQ\bQ'}\,v\,\tau_z\,\big(\bk-\bQ\big)\cdot\,\bsigma\,\\
&\qquad + \tau_0\,\sum_{i=1}^3\,\Big(\delta_{\bQ'-\bQ,\bq_i}
+ \delta_{\bQ-\bQ',\bq_i}\Big)\;\hat T_i(u,u')\,,\label{Ham-final-no-phonon}
\eal
which has the advantage of having a very compact form. For convenience,   
we list the action of $\Omega$ applied to $\sigma$ and $\tau$ operators, 
\beal
\Omega\,\sigma_x\,\Omega &= \sigma_x\,,&
\Omega\,\sigma_y\,\Omega &= \sigma_y\,\tau_z\,,&
\Omega\,\sigma_z\,\Omega &= \sigma_z\,\tau_z\,,\\
\Omega\,\tau_x\,\Omega &= \tau_x\,\sigma_x\,,&
\Omega\,\tau_y\,\Omega &= \tau_y\,\sigma_x\,,&
\Omega\,\tau_z\,\Omega &= \tau_z\,.\label{mapping-KB}
\eal
%%%%%%%%%%%%%%
In this representation, any symmetry operation $\text{G}\in D_6$ corresponds to a transformation
\beal
\hat{H}(\bk) &= \hat{D}^\dagger\big(\text{G}\big)\,  \hat{H}
\Big(\text{G}(\bk)\Big)\,\hat D\big(\text{G}\big)\,,
\eal
whose explicit expressions are given in Ref.~\cite{Bernevig_PRL}.

\section{Perturbation induced by a static atomic displacement}
\label{phonon}
We now move to derive in the continuum model the expression of the perturbation induced by 
a collective atomic displacement. 
Under a generic lattice deformation, the in-plane atomic positions $\bx_{i\alpha}$ change according to  
\beal
\bx_{i\alpha} \equiv \bR_i+\br_{i\alpha} 
&\to \bR_i+\br_\alpha + \bd{u}_i\big(\bx_{i\alpha}\big) 
= \bx_{i\alpha} + \bd{u}_i\big(\bx_{i\alpha}\big) \,,
\eal
where $i$ is now labelling a generic unit cell position.
Since the phonon modes we are going to study involve only in-plane atomic displacements, we assume that $z$-coordinate of each carbon atom does not vary.
It follows that a generic potential in the two-centres approximation  and at linear order in the displacement 
reads
\beal
&T\Big(\bx_{i\alpha}-\bx_{j\beta},
z_{i\alpha}-z_{j\beta}
\Big)\to T\Big(\bx_{i\alpha}-\bx_{j\beta},
z_{i\alpha}-z_{j\beta}\Big) \\
&\qquad\qquad + \bd{W}\Big(\bx_{i\alpha}-\bx_{j\beta},
z_{i\alpha}-z_{j\beta}\Big)\cdot
\Big(\bd{u}_i\big(\bx_{i\alpha}\big)
-\bd{u}_j\big(\bx_{j\beta}\big)\Big)
\,,
\eal
We further neglect the dependence on $z$, which we will take into account 
by distinguishing at the end between different scattering channels, intra- and inter-layers, so that:
\beal
\bd{W}(\br) &= \fract{1}{N}\,\sum_\bq\, \esp{i\bq\cdot\br}\;\bd{W}(\bq) = \bd{\nabla}\,T(\br)
= i\;\fract{1}{N}\,\sum_\bq\, \bq\;T(\bq)\,\esp{i\bq\cdot\br}\,,
\eal
namely
\beal
\bd{W}(\bq) = i\bq\,T(\bq) = i\bq\,T(q)\,,
\eal
assuming, as before, that $T(\bq)$ depends only on $q=|\bq|$. 
\subsection{ Jahn-Teller moir\'e phonon modes}

\begin{figure}
\centerline{\includegraphics[width=0.7\textwidth]{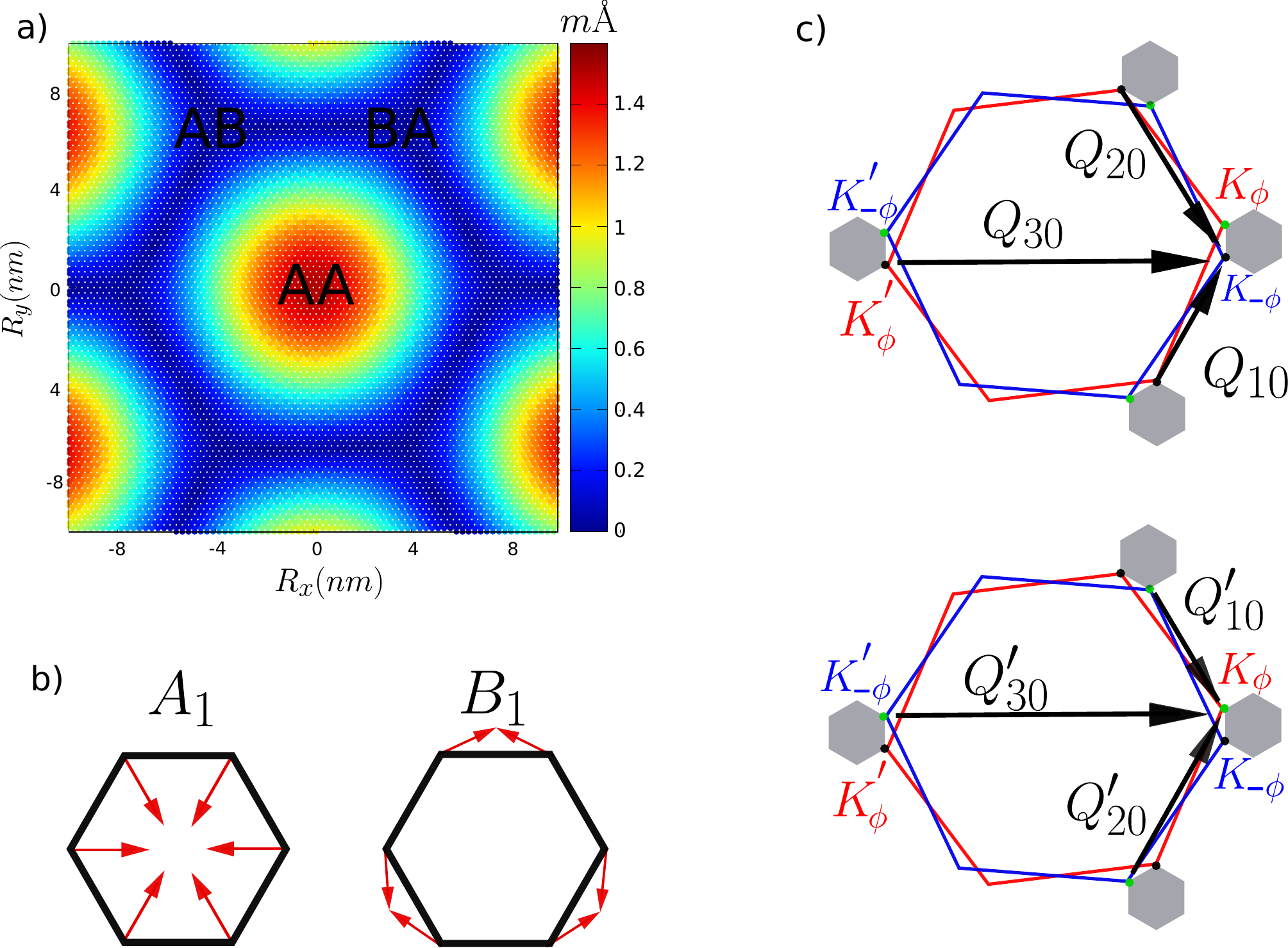}}
\caption{a) Real space phonon amplitude of the pair of Jahn-Teller phonon modes found in 
Ref.~\cite{Fabrizio_PRX}. These modes, often referred to as 'moir\'e modes', vibrate mostly within the AA regions involving no atomic movement at all within the Bernal (AB/BA) stacked regions. b) Sketch of the microscopic vibration of these modes. The atomic movement is in-plane and behave as two irreducible representations ($A_1$ and $B_1$) of the group $\text{D}_6$. c) $Q$-vectors which connect inequivalent valleys in different layers. These vectors has been used to approximate the vibration of the modes in a) and b), see \eqref{sincos}. }
\label{Fig3}
\end{figure}

Ref.~\cite{Fabrizio_PRX} pointed out the existence of a pair of high-frequency optical modes at the $\bGamma$ point of the MBZ, which are extremely efficient in lifting the valley degeneracies observed in the band structure. These phonon modes are schematically drawn in Fig.\ref{Fig3}(a), and they both share the same modulation on the moir\'e length scale. However, microscopically, they both look as the well-known in-plane optical phonon modes of graphene at $\bK$, which transform as the $A_1$ and $B_1$ irreducible representations, see Fig.\ref{Fig3}(b).
These two irreducible representations differ by the fact that $B_1$ is odd with respect to $\bC{2z}$ and $\bC{2y}$, while $A_1$ is even with respect to all symmetries of the $\text{D}_6$ space group.\\
Although the complexity of these modes is hard to capture by a simple analytical expression, their effect on the band structure can be well approximated introducing the following deformation fields
\beal
\bd{u}_c^\apa\Big(\bx^\apa_\alpha\Big) &= \sum_{i=1}^3\,\sum_{j=0}^2\, 
\bd{u}^\apa_{\alpha }\big(\bQ_{ij} \big)\,\cos \Big(\big(\bR^\apa+\br^\apa_\alpha\big)
\cdot\bQ_{ij}\Big)\,,\\
\bd{u}_s^\apa \Big(\bx^\apa_\alpha \Big) &= \sum_{i=1}^3\,\sum_{j=0}^2\, 
\bd{v}^\apa_{\alpha } \big(\bQ_{ij} \big)\,\sin \Big(\big(\bR^\apa+\br^\apa_\alpha\big)
\cdot\bQ_{ij}\Big)\,.
\label{sincos}
\eal
were $\bQ_{ij}$ are the k-vectors connecting different valleys and depicted in Fig.\ref{Fig3}(c), while  $a=1,2$ is the layer index.
Since the transformation $\bC{g}$ ($g=3z,2x$)
is a symmetry operation even in the distorted lattice, we have that  
\beal
\bC{g}\Big(\bx^\apa_\alpha\Big) &= \bx^\apb_\beta & 
&\xrightarrow{\hspace*{0.5cm}}&
\bC{g}\Big(\bd{u}^{(a)}\big(\bx^\apa_\alpha\big)\Big)
&= \bd{u}^{(b)}\big(\bx^\apb_\beta\big)\,.
\eal
By noting that the set of momenta $\{\bQ_{ij}\}$ is invariant under $\bC{3z}$ and $\bC{2x}$,  it 
immediately follows that, for $g=3z,2x$ 
\beal
\bC{g}\Big(\bd{u}^{(a)}_{\alpha }\big(\bQ_{ij}\big)\Big) &= 
\bd{u}^{(b)}_{\beta }\Big(\bC{g}\big(\bQ_{ij}\big)\Big)\,.\\
\bC{g}\Big(\bd{v}^{(a)}_{\alpha }\big(\bQ_{ij}\big)\Big) &= 
\bd{v}^{(b)}_{\beta }\Big(\bC{g}\big(\bQ_{ij}\big)\Big)\,.
\eal
On the contrary, $\{\bQ_{ij}\}$ is not invariant under $C_g$ ($g=2y,2z$), and the phonon modes are either even ($A_1$) or odd ($B_1$) under these symmetries. 
Therefore, recalling that 
 $\bC{2z}$ exchanges the two sublattices,   
\beal
\bC{2z}\Big(\bd{u}^{(a)}_{A}\big(\bQ_{ij}\big)\Big) &= - \bd{u}^{(a)}_{A}\big(\bQ_{ij}\big)
= \pm \bd{u}^{(a)}_{B}\big(\bQ_{ij}\big)\,,\\
\bC{2z}\Big(\bd{v}^{(a)}_{A}\big(\bQ_{ij}\big)\Big) &= - \bd{v}^{(a)}_{A}\big(\bQ_{ij}\big)
= \mp \bd{v}^{(a)}_{B}\big(\bQ_{ij}\big)\,.
\eal
If we choose 
\beal
\bd{u}^{(a)}_{A}\big(\bQ_{ij}\big)
&= \bd{u}^{(a)}_{B}\big(\bQ_{ij}\big)\,,&
\bd{v}^{(a)}_{A}\big(\bQ_{ij}\big)
&= \bd{v}^{(a)}_{B}\big(\bQ_{ij}\big)\,,
\eal
then the cosine distortion in \eqref{sincos} $\bd{u}_c^\apa\Big(\bx^\apa_\alpha\Big)$ transforms as $B_1$, while the sine one, $\bd{u}_s^\apa\Big(\bx^\apa_\alpha\Big)$, as $A_1$. They both can be shortly written as 
\beal
\bd{u}^\apa\Big(\bx^\apa_\alpha\Big) &= \sum_{i=1}^3\,\sum_{j=0}^2\, \Bigg[
\bd{u}^\apa\Big(\bQ_{ij}\Big)\,\esp{i\big(\bR^\apa+\br^\apa_\alpha\big)
\cdot\bQ_{ij}}\;+\,c.c.\Bigg]\,,\label{phonon-mode-expression}
\eal
where $\bd{u}^\apa\big(\bQ_{ij}\big){^*} \equiv \bd{u}^\apa\big(-\bQ_{ij}\big)$ and 
$\bd{u}^\apa\big(\bQ_{ij}\big)$ is real for the $B_1$ distortion and imaginary for $A_1$.   \\
We end by pointing out that $\bd{W}(\bq)$ satisfies
\beal
\bC{g}\big(\bd{W}(\bq)\big) &= 
\bd{W}\Big(\bC{g}\big(\bq\big)\Big)\,,
\eal
for all symmetry operations of the lattice, in particular
\beal
\bC{2z}\Big(\bd{W}(\bq)\Big) &= -\bd{W}(\bq) = 
\bd{W}\Big(\bC{2z}\big(\bq\big)\Big) = \bd{W}(-\bq)
= \bd{W}(\bq)^*
\,.%\label{W-C2z}
\eal

\subsection{Phonon induced Hamiltonian matrix elements}

A lattice distortion involving the $A_1$ or $B_1$ phonons generates a matrix element between layer $a$ momentum $\bk\sim \bK^{(a)}$ and layer $b$ momentum $\bp\sim -\bK^{(b)}$, where 
we recall that $\bK^{(1)}=\bK_2$ and $\bK^{(2)}=\bK_1$ in Fig.~\ref{fig1}(d):
\bea
W(a,\bk,\alpha;b,\bp,\beta)&=& \fract{1}{N^2}\sum_{\bq \bQ_{ij}}\sum_{\bR^\apa\bR^\apb}
\!\!\!\bd{W}(-\bq)\cdot\Bigg[\bd{u}^\apa\Big(\bQ_{ij}\Big)\,
\esp{-i\big(\bk+\bq -\bQ_{ij}\big)\cdot\big(\bR^\apa+\br^\apa_\alpha\big)}\;
\esp{i(\bp+\bq)\cdot\big(\bR^\apb+\br^\apb_\beta\big)}\nonumber\\
&&\qquad\qquad\qquad\qquad  - \bd{u}^\apb\Big(\bQ_{ij}\Big)\,\esp{-i(\bk+\bq)\cdot\big(\bR^\apa+\br^\apa_\alpha\big)}\;
 \esp{i\big(\bp +\bq +\bQ_{ij}\big)\cdot\big(\bR^\apb+\br^\apb_\beta\big)}
 \Bigg]\nonumber\\
&=& \sum_{\bq \bQ_{ij}}\sum_{\bG^\apa\bG^\apb}\!\! \bd{W}(-\bq)\cdot\Bigg[
\delta_{-\bq,\bk-\bQ_{ij}+\bG^\apa}\, \delta_{-\bq,\bp+\bG^\apb}\;
\bd{u}^\apa\Big(\bQ_{ij}\Big)\;
\esp{i\bG^\apa\cdot\br^\apa_\alpha-i\bG^\apb\cdot\br^\apb_\beta}\nonumber\\
&&\qquad\qquad\qquad\quad -
 \delta_{-\bq,\bk+\bG^\apa}\, \delta_{-\bq,\bp+\bQ_{ij}+\bG^\apb}\;
\bd{u}^\apb\Big(\bQ_{ij}\Big)\;
\esp{i\bG^\apa\cdot\br^\apa_\alpha-i\bG^\apb\cdot\br^\apb_\beta}\;\Bigg]\,.\label{W-coupling}
\eea
We can readily follow the same steps outlined in section \ref{CM} to identify 
the $\bG^\apa$ and $\bG^\apb$ reciprocal lattice vectors that enforce momentum 
conservation and maximise the matrix element $W(-\bq)=W(q)$. Therefore, we shall not repeat 
that calculation and jump directly to the results. \\

\noindent
The lattice distortion introduces a perturbation both intra-layer and  
inter-layer. The former, in the representation introduced in Sect.~\ref{A more convenient representation}, has the extremely simple expression:
\beal
\delta \hat H_{x(y)}^{||}(\bk){_{\bQ\bQ'}} &=  
\tau_{x(y)}\,\sum_{i=1}^3\,\Big(\delta_{\bQ'-\bQ,\bq_i}
+ \delta_{\bQ-\bQ',\bq_i}\Big)\;\hat T_i(g,g')
\equiv \tau_{x(y)}\;\delta \hat H^{||}_{\bQ\bQ'}
\,,\label{ph-intra}
\eal
where $\tau_x$ refers to the $A_1$ mode, $\tau_y$ to the $B_1$ one, 
and the matrices $\hat T_i(g,g')$ have the same expression as those in 
Eq.~\eqn{relax}, with $u$ and $u'$ replaced, respectively, by $g$ and $g'$. \\

\noindent
The inter layer coupling has a simpler expression, since, as we many times mentioned, 
the opposite valleys in different layers fold on the same momentum in the MBZ, and thus 
the coupling is diagonal in $\bQ$ and $\bQ'$ and reads
\beal
\delta \hat H_{x(y)}^{\perp}(\bk)_{\bQ\bQ'} &= \delta_{\bQ,\bQ'}\;\gamma\,\sigma_0\,\tau_{x(y)}
\equiv \tau_{x(y)}\;\delta \hat H^{\perp}_{\bQ\bQ'}\,.
\label{ph-inter}
\eal
As before $\tau_x$ and $\tau_y$ refers to the $A_1$ and $B_1$ modes, respectively.\\
It is worth remarking that, because of the transformation \eqn{U_oper}, 
which exchanges the sublattices in the valley $\zeta=-1$,  
the diagonal elements of the matrices $\hat T_i(g,g')$ in \eqn{ph-intra} 
and $\sigma_0$ in \eqn{ph-inter} refer to the opposite sublattices, while 
the diagonal elements to the same sublattice, right the opposite of the unperturbed 
Hamiltonian \eqn{Ham-final-no-phonon}. \\

\noindent
Let us rephrase the above results in second quantisation and introducing the 
quantum mechanical character of the phonon mode. In the continuum model, a plane wave 
with momentum $\bk+\bG$, where $\bG=n\bd{A}+m\bd{B}$ is a reciprocal lattice vector of the 
MBZ, in layer $i=1,2$, valley $\zeta=+1$ and with sublattice components described 
by a two-component spinor $\chi_{\bk+\bG}$ can be associated to a two component spinor operator
according to 
\beal
\esp{i(\bk+\bG)\cdot\br}\;\chi_{\bk+\bG}\; \xrightarrow{\hspace*{0.5cm}}\;  \Psi^{(i)}_{+1,\bk+\bG}\,.
\eal
For any $\bG$ I can write, see Eq.~\eqn{def:Q-vectors},
\beal
\bG &= \bK_1 -\bQ_A = \bK_2-\bQ_B\,,
\eal
and thus define 
\beal
\Psi^{(1)}_{+1,\bk+\bG} &= \Psi^{(1)}_{+1,\bk+\bK_{2}-\bQ_B} \equiv \Psi_{\bk,\bQ_B,+1}\,,\\
\Psi^{(2)}_{+1,\bk+\bG} &= \Psi^{(1)}_{+1,\bk+\bK_{1}-\bQ_A} \equiv \Psi_{\bk,\bQ_A,+1}\,.
\eal
I note that $\bK_{+\phi}+\bK_{-\phi}\equiv\bG_k = (2k+1)\,\big(\bd{A}+\bd{B}\big)$, 
which allows us defining the operators in valley $\zeta=-1$ as
\beal
\Psi^{(1)}_{-1,\bk+\bG-\bG_k} &= \Psi^{(1)}_{-1,\bk-\bK_{+\phi}-\bQ_A} \equiv \sigma_x\,\Psi_{\bk,\bQ_A,-1}\,,\\
\Psi^{(2)}_{-1,\bk+\bG-\bG_k} &= \Psi^{(2)}_{+1,\bk-\bK_{-\phi}-\bQ_B} \equiv 
\sigma_x\,\Psi_{\bk,\bQ_B,-1}\,,
\eal
where, in accordance with our transformation in Eq.~\eqn{U_oper}, we interchange the 
two sublattices in valley $\zeta=-1$ through $\sigma_x$.
We note that the mismatch momentum $\bG_k$ is just what is provided by the phonon modes. Absorbing the 
valley index into two additional components of the spinors, and introducing back the spin label, the second quantised Hamiltonian can be written in terms of four 
component spinor operators $\Psi_{\bk\sigma,\bQ}$, where, $\bQ=\bQ_A$ 
refer to layer 2 if the valley index $\zeta=+1$ and layer 1 if $\zeta=-1$, 
while $\bQ=\bQ_B$ to layer 1 if $\zeta=+1$, and layer 2 if $\zeta=-1$.\\  
Next, we introduce a two component dimensionless variable $\bq_\bd{0}=(q_1, q_2)$, and its conjugate one, 
$\bp_\bd{0}=(p_1, p_2)$, where $q_1$ and $q_2$ are the phonon coordinates of the 
$A_1$ and $B_1$ modes at $\bGamma=\bd{0}$, respectively. Using the above defined operators,  
the full quantum mechanical Hamiltonian reads  
\beal
H &= \sum_{\bk\bQ\sigma}\, \bigg[\;\Psi^\dagger_{\bk\sigma\,\bQ}\;\hat H_{\bQ\bQ'}(\bk)\;\Psi^\dagga_{\bk\sigma\,\bQ'} + \Psi^\dagger_{\bk\sigma\,\bQ}
\Big( \bd{q}_\bd{0}\cdot\bd{\tau}\;\delta \hat H_{\bQ\bQ'} \Big)
\,\Psi^\dagga_{\bk\sigma\,\bQ'}\;\bigg]\\
&\qquad\qquad \qquad
+ \fract{\omega_\bd{0}}{2}\,\Big(\,\bp_\bd{0}\cdot\bp_\bd{0} + \bq_\bd{0}\cdot\bq_\bd{0}\Big)\,,\label{Ham-QM}
\eal
where $\omega_\bd{0}$ is the phonon frequency, equal for both $A_1$ and $B_1$ modes, 
$\bd{\tau}=(\tau_x,\tau_y)$, 
and 
\beal
\hat{H}_{\bQ \bQ'}(\bk)  &= \delta_{\bQ\bQ'}\,v\,\tau_z\,\big(\bk-\bQ\big)\cdot\,\bsigma
+ \tau_0\,\sum_{i=1}^3\,\Big(\delta_{\bQ'-\bQ,\bq_i}
+ \delta_{\bQ-\bQ',\bq_i}\Big)\;\hat T_i(u,u')\,,\\
\delta \hat H_{\bQ\bQ'} &= \delta \hat H^{||}_{\bQ\bQ'} + \delta \hat H^{\perp}_{\bQ\bQ'}
= \sum_{i=1}^3\,\Big(\delta_{\bQ'-\bQ,\bq_i}
+ \delta_{\bQ-\bQ',\bq_i}\Big)\;\hat T_i(g,g')
+ \delta_{\bQ,\bQ'}\;\gamma\,.\label{delta-H-final}
\eal
We observe that the Hamiltonian 
\eqn{Ham-QM} still possesses a valley $U_v(1)$ symmetry, with generator
\beal
J_z &= \fract{1}{2}\,\sum_{\bk\bQ\sigma}\,\Psi^\dagger_{\bk\sigma\,\bQ}\;\sigma_0\,\tau_z
\;\Psi^\dagga_{\bk\sigma\,\bQ} + \bq_\bd{0}\wedge\bp_\bd{0} \equiv T_z + L_z\,,
\eal
where $T_z$ is half the difference between the number of electrons in valley 
$\zeta=+1$ and the one in valley $\zeta=-1$, while $L_z$ is the angular momentum of the phonon mode. 
The Hamiltonian \eqn{Ham-QM} actually realises a $e\otimes E$ Jahn-Teller model.\\

\noindent
\begin{figure}
\centerline{\includegraphics[width=1.05\textwidth]{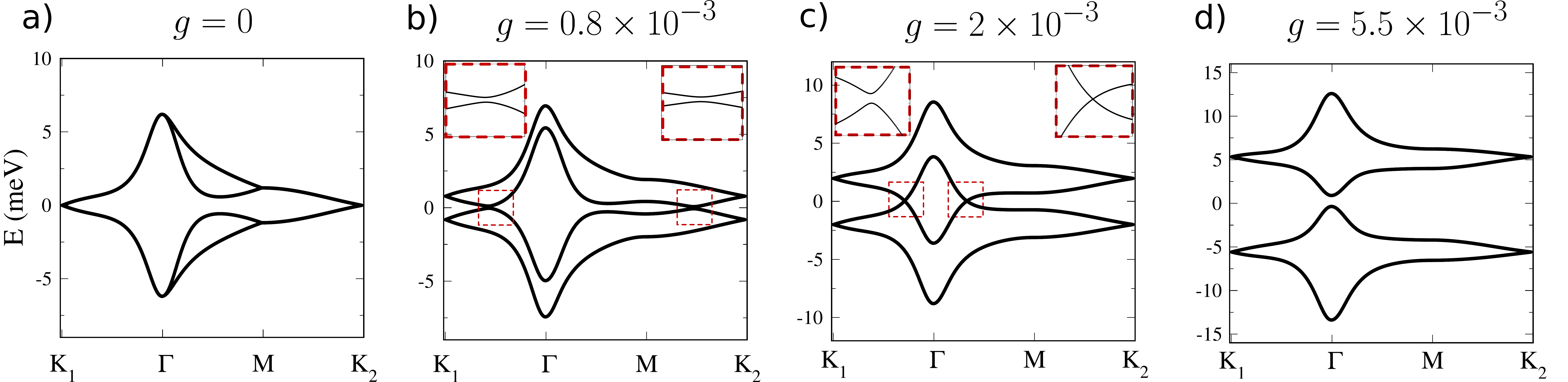}}
\caption{Band structure of $2\phi=1.08^\circ$ 
twisted bilayer graphene with increasing frozen phonon deformation intensity $g$. The other parameters used in this calculation are: $u=0.0761$, $u'=0.1031$, $g'=g/10$ and $\gamma=g/2.5$. a) Undistorted case. b) Slightly distorted lattice. Two small avoided crossings occur along the $\bK_1 \to \bGamma$ and $\bM\to \bGamma$ lines. c) By further increasing the distortion intensity the four bands further separate. The avoided crossing along $\bK_1 \to \bGamma$ persists, while a genuine crossing occur along $\bGamma \to \bM$. d) A gap at charge neutrality has finally opened.}
\label{bands}
\end{figure}

It is straightforward to generalise the above result to an atomic displacement modulated with the wave vectors $\bQ_{ij} + \bd{P}$, where  $\bd{P}\in\text{MBZ}$. Since $\bQ_{ij}$ are multiples of the MBZ reciprocal lattice vector, such displacement is at momentum $\bd{P}$, and can be considered as the previous one at $\bGamma$, shown in Fig.~\ref{Fig3}, on top of which we add an additional incommensurate long wavelength component.  Since $\bd{P}$ is tiny as compared to the vectors $\bQ_{ij}$, we shall assume that the displacement has the same expression of Eq.~\eqn{phonon-mode-expression}, with the only difference that 
\beal
\esp{\pm i\big(\bR^\apa+\br^\apa_\alpha\big)
\cdot\bQ_{ij}} \; &\xrightarrow{\hspace*{0.6cm}}\;
\esp{\pm i\big(\bR^\apa+\br^\apa_\alpha\big)
\cdot\big(\bQ_{ij}\pm \bd{P}\big)} \;. \label{simplification}
\eal
The full quantum mechanical Hamiltonian becomes 
\beal
H &= \sum_{\bk\bQ\sigma}\, \Psi^\dagger_{\bk\sigma\,\bQ}\;\hat H_{\bQ\bQ'}(\bk)\;\Psi^\dagga_{\bk\sigma\,\bQ'} + \sum_{\bk\bQ\bd{P}\sigma}\, \Psi^\dagger_{\bk\sigma\,\bQ}
\Big( \bd{q}_{-\bd{P}}\cdot\bd{\tau}\;\delta \hat H_{\bQ\bQ'}(\bd{P}) \Big)
\,\Psi^\dagga_{\bk+\bd{P}\sigma\,\bQ'}\\
&\qquad\qquad \qquad
+ \fract{1}{2}\,\sum_{\bd{P}}\, \omega_\bd{P}\,\Big(\,\bp_\bd{P}\cdot\bp_{-\bd{P}} + \bq_\bd{P}\cdot\bq_{-\bd{P}}\Big)\,,\label{Ham-QM-final}
\eal
where $\delta \hat H_{\bQ\bQ'}(\bd{P})$ is the same as $\delta \hat H_{\bQ\bQ'}$ in Eq.~\eqn{delta-H-final} with $\bd{P}$-dependent constants $g_\bd{P}$, $g'_\bd{P}$, and $\gamma_\bd{P}$, invariant 
under the little group at $\bd{P}$.  In this general case, the generator of  $U_v(1)$ reads
\beal
J_z &= \fract{1}{2}\,\sum_{\bk\bQ\sigma}\,\Psi^\dagger_{\bk\sigma\,\bQ}\;\sigma_0\,\tau_z
\;\Psi^\dagga_{\bk\sigma\,\bQ} + \sum_{\bd{P}}\,\bq_\bd{P}\wedge\bp_{-\bd{P}} \,.
\eal

\subsection{Frozen phonon band structure}

We can perform a frozen phonon calculation neglecting the phonon energy, last term in Eq.~\eqn{Ham-QM}, and fixing $\bq=(q_1,q_2)$ to some constant value. Because of the $U_v(1)$ symmetry, what matters 
is just the modulus $q$ of $\bq$. In practice we have taken $\bq=(1,0)$, and studied the band structure 
varying the coupling constants $g$, setting $g'=g/10$ and $\gamma=g/2.5$, and 
assuming the following parameters: $\hbar v/a_0=2.1354\;eV$ \cite{Koshino_PRB};
$u=0.0761\; eV$ and $u'=0.1031 \;eV$ \cite{Procolo}. This choice fits well 
the microscopic tight-binding calculations in Ref.~\cite{Fabrizio_PRX}. 
As shown in Fig.\ref{bands}, as soon as the frozen phonon terms are turned on, all the degeneracies in the band structure arising due to the valley symmetry are lifted. This occur with a set of avoided crossings which move from $\bK \to \bM$ and from $\bK \to \bGamma$ (Fig.\ref{bands}b)). In particular, the crossings that move from $\bK \to \bM$ eventually meet at $\bM$, forming (six) Dirac nodes, which then move towards $\bGamma$ (Fig.\ref{bands}c)). Finally, at a threshold value of $g$, a gap opens at the charge neutrality point (Fig.\ref{bands}d)). Such gap keeps increasing as the deformation amplitude increases.

\subsection{Moir\'e phonons at $\bM$}

\begin{figure}
%\vspace{-1cm}
\centerline{\includegraphics[width=0.6\textwidth]{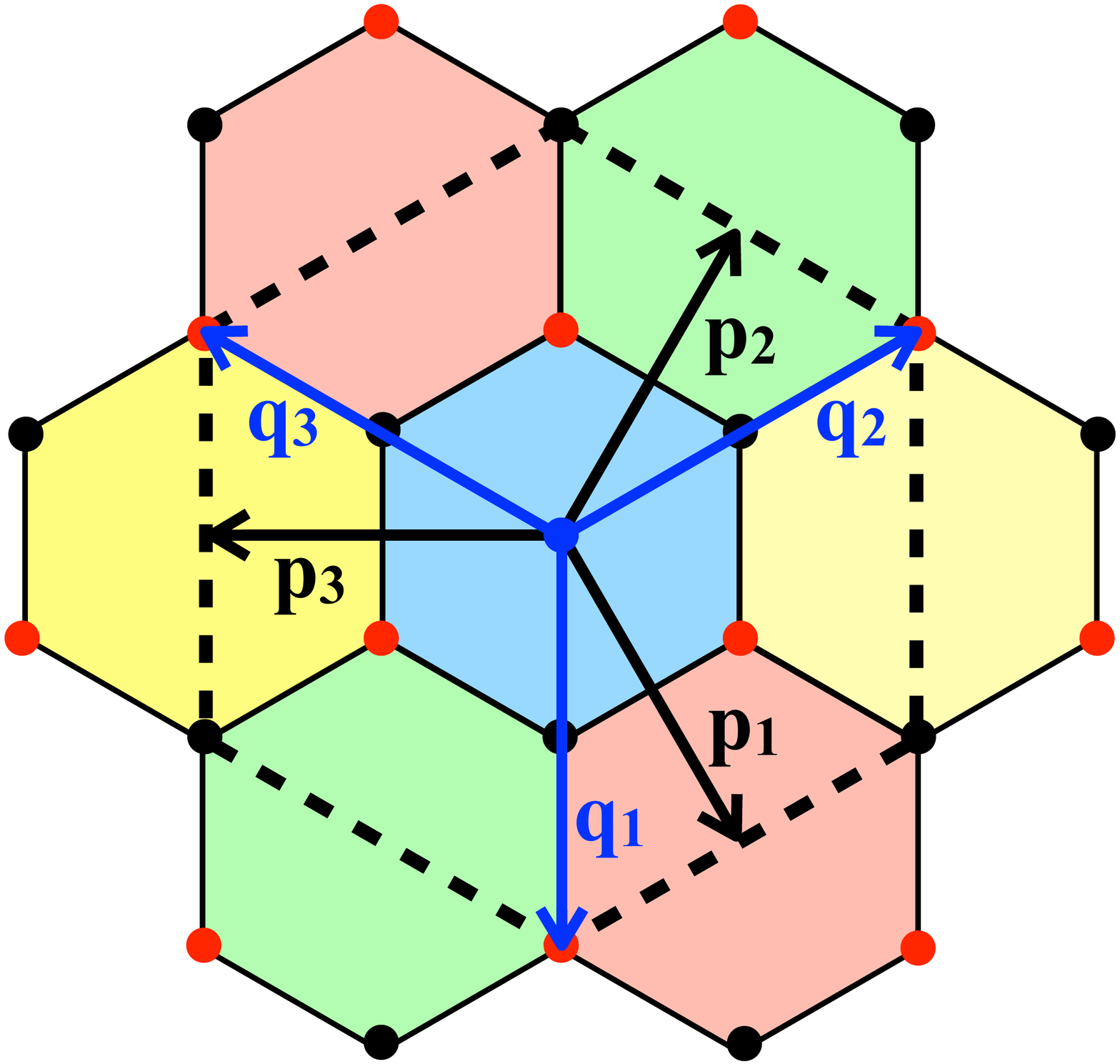}}
%\vspace{-.5cm}
\caption{Reduced Brillouin zones with the multicomponent distortion at $\bM$, shown as coloured hexagons, while the four times larger hexagon drawn with a dashed line is the undistorted MBZ. The vectors $\bp_i$, black arrows, and $\bq_i$, blue arrow, for $i=1,\dots,3$, are also shown. Note that $\bp_1=\bd{A}'$ and $\bp_2=\bd{B}'$ are also the reciprocal lattice vectors of the reduced Brillouin zone. The black and red circles are the positions of the sublattice vectors $\bQ_A$ and $\bQ_B$, respectively, defined in Eq.~\eqn{def:Q-vectors-M}.}\label{3M}
\end{figure}

The phonon modes considered in the previous section were at position $\bGamma$ of the MBZ, thus preserving the periodicity of the moir\'e superlattice. 
As pointed out in Ref.~\cite{Fabrizio_PRX} and shown before, these modes are able to open a gap in the band structure only at charge neutrality. Gap opening at different commensurate fillings requires freezing finite momentum phonons \cite{Fabrizio_PRX}.
Here, we consider a multicomponent distortion which involves the modes at the three inequivalent $\bM$ points in the MBZ:
\beal
\bM_1 &= \fract{\bd{A}}{2}\;,&
\bM_2 &= \bC{3z}\big(\bM_1 \big)= \fract{\bd{B}}{2}\;,&
\bM_3 &= \bC{3z} \big(\bM_2 \big) = -\fract{\bd{A}+\bd{B}}{2}\;.
\eal
Freezing a multiple distortion at all these points reduces by a quarter the Brillouin zone, see 
Fig.~\ref{3M}, which has now the reciprocal lattice vectors
\beal
\bd{A}' &= \bM_1
\;,& \bd{B}' &= \bM_2\;.
\eal

Since $\bM_i$, $i=1,2,3$, are tiny as compared to the vectors $\bQ_{ij}$ introduced 
in the previous section, we can make the same assumption \eqn{simplification} that 
leads to the Hamiltonian \eqn{Ham-QM-final}, namely  assume that the displacement induced by the multiple distortion has the same expression of Eq.~\eqn{phonon-mode-expression}, with the only difference that 
\beal
\esp{i\big(\bR^\apa+\br^\apa_\alpha\big)
\cdot\bQ_{ij}} \; &\xrightarrow{\hspace*{0.6cm}}\;
\sum_{n=1}^6\,\esp{i\big(\bR^\apa+\br^\apa_\alpha\big)
\cdot\big(\bQ_{ij}+\bp_n\big)}
\eal
where 
\beal
\bp_1 &=-\bp_4=\bd{M}_1=\bd{A}'\,,&
\bp_2 &= -\bp_5 = \bd{M}_2=\bd{B}'\,,& 
\bp_3 &=-\bp_6=\bd{M}_3=-\bd{A}'-\bd{B}'\,,
\eal
are the additional long wavelength modulation 
vectors on top of the leading short wavelength ones at $\bQ_{ij}$. 
\begin{figure}
%\vspace{-1cm}
\centerline{\includegraphics[width=1.1\textwidth]{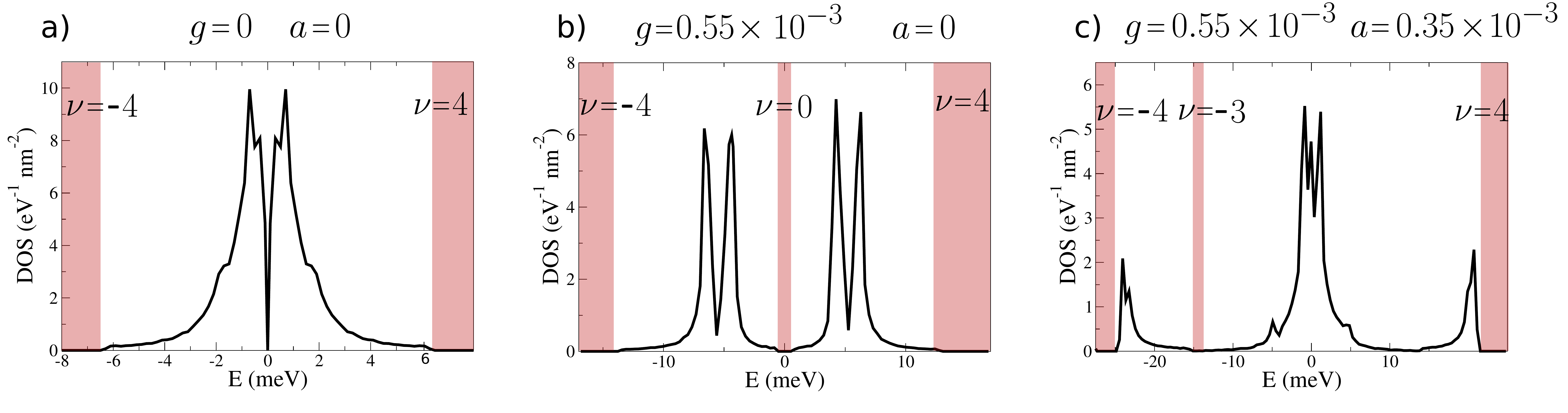}}
%\vspace{-.5cm}
\caption{ Density of states around charge neutrality obtained with the Hamiltonian \eqref{Ham-phon-M} with $g'=g/10$, $\gamma=g/2.5$, $a'=a/10$ and $\alpha=a/2.5$. Gaps are highlighted in red, and the corresponding filling factor is $\nu$.  a) The undistorted lattice density of states. b) The density of states obtained by deforming the lattice with only the distortion at $\bGamma$, which opens a gap at charge neutrality. c) Density of states obtained using both the $\Gamma$ and the $\bM$ multicomponent distortions. Here a gap opens at filling of 1 electron (3 holes with respect to neutrality) per unit cell.}\label{DOS}
\end{figure}
The vectors $\bq_i$ defined in Eq.~\eqn{def:q} can be written in terms of the new reciprocal 
lattice vectors $\bd{A}'$ and $\bd{B}'$ as   
\beal
\bq_1 &= 
\fract{2}{3}\;\big(\bd{A}'-\bd{B}'\big) 
\,,&
\bq_2 &= \fract{2}{3}\;\big(\bd{A}'+2\bd{B}'\big)\,,&
\bq_3 &= \fract{2}{3}\;\big(-2\bd{A}'-\bd{B}'\big)
\,.
\eal
Both $\bp_i$ and $\bq_i$, $i=1,2,3$, are shown in Fig.~\ref{3M}. 
Considering all momenta $\bk$ within the new BZ, the light blue hexagon in Fig.~\ref{3M}, and assuming that, besides the multicomponent distortion at $\bM$, there is still a distortion at $\bGamma$, 
the Hamiltonian can be written again as a matrix $\hat{H}_{\bQ\bQ'}(\bk)$, which now reads
\bea
\label{Ham-phon-M}
\hat{H}_{\bQ\bQ'}(\bk) &=& \delta_{\bQ,\bQ'}\; v\,\tau_z\,\Big(\bk-\bQ\Big)\cdot\bsigma
 + \tau_0\,\sum_{i=1}^3\, \Big(
\delta_{\bQ'-\bQ,\bq_i} + \delta_{\bQ-\bQ',\bq_i}\Big)\;\hat{T}_i\big(u,u')\nonumber\\
&&\qquad + \gamma\,\delta_{\bQ,\bQ'}\;\tau_x 
+ \tau_x\,\sum_{i=1}^3\, \Big(
\delta_{\bQ'-\bQ,\bq_i} + \delta_{\bQ-\bQ',\bq_i}\Big)\;\hat{T}_i\big(g,g')\\
&&\; +\alpha\,\tau_x\sum_{i=1}^6 \Big(
\delta_{\bQ'-\bQ,\bp_i} + \delta_{\bQ-\bQ',\bp_i}\Big)
+ \tau_x\,\sum_{i=1}^3 \sum_{j=1}^6\Big(
\delta_{\bQ'-\bQ,\bq_i+\bp_j} + \delta_{\bQ-\bQ',\bq_i+\bp_j}\Big)\;\hat{T}_i\big(a,a')\,, 
\nonumber
\eea
where the matrices $\hat T_i(x,x')$ are those in Eq.~\eqn{relax}, though they depend on different set 
of parameters, $(u,u')$,  $(g,g')$ and $(a,a')$.  
The crucial difference with respect to the Hamiltonian \eqn{delta-H-final} 
with only the $\bGamma$-distortion, is that the $\bQ$ vectors span now the sites of 
the honeycomb lattice generated by the new fourfold-smaller Brillouin zone, hence they are defined through 
\beal
\bQ &= \Big\{\bQ'_A\,,\,\bQ'_B\Big\} =
\begin{cases}
\bQ'_A = \fract{\bd{A}'-\bd{B}'}{3} + n\bd{A}'+m\bd{B}'\,,\\
\bQ'_B = -\fract{\bd{A}'-\bd{B}'}{3} + n\bd{A}'+m\bd{B}'\,,
\end{cases}\label{def:Q-vectors-M}
\eal
and shown in Fig.~\ref{3M} as black and red circles, respectively, and must not be confused with those in Eq.~\eqn{def:Q-vectors}. 
In Fig.~\ref{DOS} we show the density of states around neutrality of the Hamiltonian \eqref{Ham-phon-M}. The first two cases corresponds to undistorted and $\Gamma$-only distorted structures, while the third panel involves also the $M$ multicomponent distortion. As can be seen, a gap now opens at the partial filling of 1 electron per unit cell. As it was shown in Ref.~\cite{Fabrizio_PRX}, other phonons or combinations of them can open gaps at any integer filling of the four electronic flat bands.

\section{Conclusions}
\label{conclusion}
We have shown that the \mo phonons of Ref.~\cite{Fabrizio_PRX}, which are coupled 
to the valley degrees of freedom of the electrons so to realise an $E\otimes e$ Jahn-Teller 
model, can be successfully implemented in the continuum model formalism of small angle twisted bilayer graphene. This method is more manageable than the realistic tight-binding modelling of Ref.~\cite{Fabrizio_PRX}, whose results have been here reproduced with much less 
effort. In addition, the continuum model formalism has the great advantage of providing a full quantum mechanical expression of the electron-phonon Hamiltonian, which 
may allow going beyond the simple frozen-phonon calculation of \cite{Fabrizio_PRX}, 
and thus describing phenomena like a dynamical Jahn-Teller effect and the phonon-mediated 
superconductivity. 

\section*{Acknowledgments}
We acknowledge useful discussions with A.H.~MacDonald. This work has been supported by the European Research Council (ERC) under H2020 Advanced Grant No. 692670 ``FIRSTORM''.

%
% For  figures use
%\begin{figure*}
%\vspace*{5cm}       % Give the correct figure height in cm
%\includegraphics{leer.eps}
%\end{figure*}
% or  this
%\begin{figure}
%\centering
%\resizebox{0.75\textwidth}{!}{%
%  \includegraphics{leer.eps}
%\vspace{5cm}       % Give the correct figure height in cm
%\end{figure}
%\begin{table}
%\centering
%\begin{tabular}{lll}
%\hline\noalign{\smallskip}
%\end{tabular}
% Or use
%\vspace*{5cm}  % with the correct table height
%\end{table}

%
% BibTeX users please use
 \bibliographystyle{abbrv}

% \bibliography{mybiblio.bib}
%

\end{document}